\documentclass[aps,prb,reprint,floatfix,longbibliography,twocolumn]{revtex4-2}
\usepackage{amsmath,amssymb,amsthm,hyperref,graphicx}
\hypersetup{hidelinks}
\usepackage[capitalise]{cleveref}
\hypersetup{colorlinks=true, linkcolor=blue, citecolor=red, urlcolor=blue}
\begin{document}
\title{Strong and tunable coupling between antiferromagnetic magnons and surface plasmons}
\author{H. Y. Yuan$^{1}$}
\email{Contact author: hyyuan@zju.edu.cn}
\author{Yaroslav M. Blanter$^{2}$}
\author{H. Q. Lin$^{1}$}
\affiliation{$^{1}$Institute for Advanced Study in Physics, Zhejiang University, 310027 Hangzhou, China}
\affiliation{$^{2}$Department of Quantum Nanoscience, Kavli Institute of Nanoscience, Delft University of Technology, 2628 CJ Delft, The Netherlands}


\date{\today}

\begin{abstract}
Surface plasmons are the collective electron excitations in metallic systems and the associated electromagnetic wave usually has the transverse magnetic (TM) polarization. On the other hand, spin waves are the spin excitations perpendicular to the equilibrium magnetization and are usually circularly polarized in a ferromagnet. The direct coupling of these two modes is difficult due to the difficulty of matching electromagnetic boundary conditions at the interface of magnetic and non-magnetic materials. Here, we overcome this challenge by utilizing the linearly polarized spin waves in antiferromagnets (AFM) and show that a strong coupling between AFM magnons and surface plasmons can be realized in a hybrid 2D material/AFM structure, featuring a clear anticrossing spectrum at resonance. The coupling strength, characterized by the gap of anticrossing at resonance, can be tuned by electric gating on 2D materials and be probed by measuring the two reflection minima in the reflection spectrum. Further, as a potential application, we show that plasmonic modes can assist the coupling of two well-separated AFMs over several micrometers, featuring symmetric and antisymmetric hybrid modes. Our results may open a new platform to study antiferromagnetic spintronics and its interplay with plasmonic photonics.
\end{abstract}

\maketitle
\section{Introduction}
Understanding the light-matter interaction is a central topic in condensed matter physics. Light can induce electron oscillations on a metallic surface and in 2D conducting materials, so-called surface plasmons \cite{Ritchie1957, SternPR1960, Meierbook}, which is promising if one wants to confine and amplify electromagnetic waves for sensing applications \cite{ZayatsPR2005, RodrigoScience2015,IranzoScience2018}. While surface plasmons in the optical and infrared regime have been widely studied, their extension down to gigahertz (GHz) and terahertz (THz) regime with desirable polarity has been an outstanding challenge for a long time. This is because metals behave as perfect conductors in the low- frequency regime and almost reflect all the incident electromagnetic waves, hindering the formation of surface plasmons. With the development of fabrication technology, one can grow artificially structured materials and 2D materials to generate THz spoof surface plasmons with similar dispersion and sub-wavelength field confinement compared to the conventional surface plasmons \cite{PendryScience2004, GrigorenkoNP2012}.

On the other hand, besides the charge degree of freedom, electrons also have intrinsic spin. Light can stimulate collective excitations of exchange-coupled localized spins in magnetic materials, so-called spin waves \cite{Battiato2010,Lich2022}. The quasiparticles corresponding to spin waves are magnons. Investigating spin-wave transport and novel magnon states and their potential applications in information processing is the focus of magnon spintronics \cite{ChumakNP2015,YuanQM}. It has been shown that, by combining the 2D materials such as graphene with magnetic systems, one can generate a transverse electric (TE) surface plasmon ranging from GHz to THz regime under the assistance of surface spin waves \cite{Yuanarxiv2024}. Such a proposal is free from the constrains of the conductivity of 2D materials and further benefits the great tunability of magnetic systems and 2D materials by external magnetic fields and electric gating, respectively. Nevertheless, to achieve coherent and reliable information transfer between two parties, such as magnon-photon and magnon-qubit system, it is desirable to realize strong coupling between them \cite{YuanQM,Rameshti2022}. This is not easy in layered structures involving ferromagnets and 2D materials, because the spin oscillations in a ferromagnet are circularly polarized, and it is challenging to guarantee the continuity of electromagnetic boundary conditions at the interface.

Here, we take a step further to show that surface plasmons can strongly couple to surface magnons in antiferromagnets (AFM), featuring a typical anticrossing spectrum. The essential physics is that left and right circularly polarized spin waves coexist in an AFM, and their superposition could generate a linearly-polarized spin wave. A transverse magnetic wave incident on the interface of 2D materials and AFM could be efficiently coupled to the linearized polarized spin wave and further enhance the plasmon excitation in 2D material. This is a unique feature of AFM, and is absent in ferromagnetic thin films, which only allows the excitation of circularly polarized spin wave. The THz nature of magnon frequency in an AFM enables the excitation of THz plasmons, with a reduced wavelength compared to the vacuum light. To probe the excitation, we propose to measure the reflection spectrum of the hybrid system, where the hybrid magnon-plasmon excitations carry away electromagnetic energy and generate a double-valley structure. As a potential application, we show that plasmons can mediate the coupling of two AFMs separated over several micrometers, even though the AFMs have vanishingly small magnetization. Our work may open a novel route for the interdisciplinary development of antiferromagnetic spintronics and plasmonic photonics.

\begin{figure}
	\centering
	\includegraphics[width=0.48\textwidth]{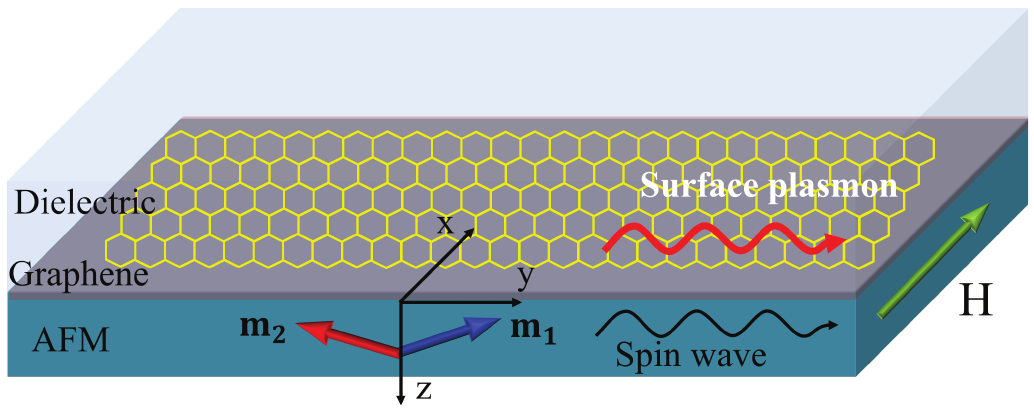}\\
	\caption{Scheme of a dielectric/graphene/AFM structure. The surface spin waves excited in the AFM is strongly coupled to the transverse magnetic surface plasmons excited in the graphene layer. The blue and red arrows represent the sublattice magnetizations of the antiferromagnetic layer.}\label{fig1}
\end{figure}


\section{Model and physical picture} 
We consider a dielectric/graphene/antiferromagnetic insulator (DE/GRA/AFM) hybrid structure as shown in Fig. \ref{fig1}.  The AFM layer constitutes two magnetic sublattices, and the magnetization dynamics is described by the two coupled Landau-Lifshitz-Gilbert (LLG) equations
\begin{subequations}\label{twoLLG}
\begin{align}
\partial_t \mathbf{m}_1 =-\gamma \mathbf{m}_1 \times \mathbf{H}_{\mathrm{1,eff}} + \alpha \mathbf{m}_1 \times \partial_t \mathbf{m}_1,\\
\partial_t \mathbf{m}_2 =-\gamma \mathbf{m}_2 \times \mathbf{H}_{\mathrm{2,eff}} + \alpha \mathbf{m}_2 \times \partial_t \mathbf{m}_2,
\end{align}
\end{subequations}
where $\gamma$ is the gyromagnetic ratio, $\mathbf{M}_i = M_s \mathbf{m}_i$ with $M_s$ being the saturation magnetization of each sublattice and $\mathbf{m}_i$ being a unit vector representing the direction of magnetization, $\alpha$ is Gilbert damping coefficient characterizing the relaxation rate of magnetization. The effective fields $\mathbf{H}_{i,\mathrm{eff}}~(i=1,2)$ include the contributions from the exchange field, anisotropy field, external field ($\mathbf{H}_e$) and dipolar field ($\mathbf{H}_d$), i.e.
\begin{subequations}
\begin{align}
\mathbf{H}_{\mathrm{1,eff}}=-H_\mathrm{ex} \mathbf{m}_2 + H_\mathrm{an} m_{1,x}e_x + \mathbf{H}_e + \mathbf{H}_d,\\
\mathbf{H}_{\mathrm{2,eff}}=-H_\mathrm{ex} \mathbf{m}_1 + H_\mathrm{an} m_{2,x}e_x + \mathbf{H}_e + \mathbf{H}_d.
\end{align}
\end{subequations}

\begin{figure}
	\centering
	\includegraphics[width=0.48\textwidth]{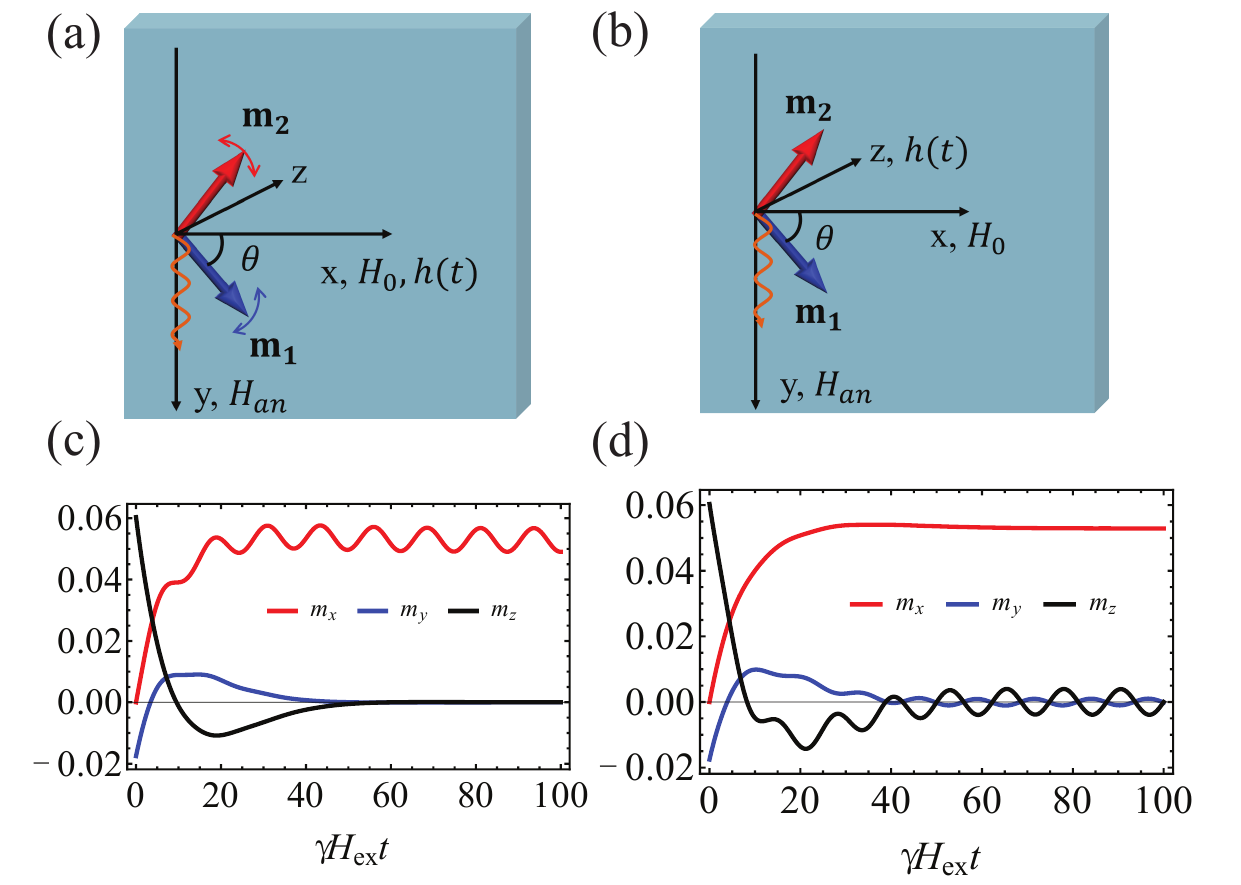}\\
	\caption{Scheme of the two sublattice antiferromagnet under a perpendicular magnetic field with driving field parallel (a) and perpendicular (b) to the static magnetic field ($x-$axis). (c-d) Time evolution of the total magnetization $\mathbf{m}=\mathbf{m}_{1} + \mathbf{m}_{2}$. Magnetic parameters of NiO are used with $H_\mathrm{ex}=524$ T, $H_\mathrm{an} = 1.47$ T, $M_s=0.32$ T \cite{YuanPRR2019}, $\gamma H_0=0.5\omega_\mathrm{sp}$. $\alpha$ is taken to be 0.1 to accelerate the relaxation process.}\label{fig2}
\end{figure}

The classical ground state of the system in the absence of external magnetic fields is a N\'{e}el state with $\mathbf{M}_1=M_se_y,~\mathbf{M}_2=-M_se_y$. When an external field is applied perpendicular to the easy axis ($y-$axis), i.e. $\mathbf{H}_e=H_0e_x$ the ground state becomes a tilted state as shown in Fig. \ref{fig2}(a-b) with the tilting angle $\cos \theta = H_0/(2H_\mathrm{ex}+H_\mathrm{an})$. Depending on the direction of the oscillating field $h(t)$, two excitation modes of magnetization can be generated. When the driving field is applied along the $z-$axis, both $m_{y}$ and $m_{z}$ oscillate with time as shown in Fig. \ref{fig2}(d). Such a geometry can help to excite the TE surface plasmons by matching the electromagnetic boundary conditions at the interface of the 2D material and the magnet, as verified in Ref. \cite{Yuanarxiv2024}. However, there is no anticrossing between magnons and plasmons due to the absence of bare TE plasmon mode. 

On the other hand, when the oscillating field is applied along the direction of static magnetic field ($x-$axis), one can have a linearly-polarized spin wave with only $m_{x}$ oscillating in time, as shown in Fig. \ref{fig2}(c). Considering the spin wave propagating in the $y$-axis, such an oscillation will generate a TM spin wave and thus can match the input TM electromagnetic wave to excite TM surface plasmons at the 2D layer, as we shall see below. Since a bare TM plasmon is not forbidden in the current geometry, one can expect an anticrossing structure when the frequencies of the surface spin wave and the surface plasmon are close. This is the main idea of our proposal. Next, we shall rigorously verify this point.

\section{Hybrid magnon-plasmon excitation} 
To study the spin-wave excitation above the ground state shown in Fig. \ref{fig2}(a), we do linear expansions of the sublattice magnetization around the tilted state as $\mathbf{M}_1=M_s \cos \theta e_x + M_{1,x}e_x + M_{1,y}e_y,~\mathbf{M}_2=-M_s \cos \theta e_x + M_{2,x}e_x + M_{2,y}e_y$. Substituting this trial solution into the coupled LLG equations \eqref{twoLLG} and keeping only the linear terms, we derive that 
$M_{1,y}+ M_{2,y}=0$ and 
\begin{equation}
\begin{aligned}
\partial_{tt} \delta {M_x}&=-\omega^2_\mathrm{sp} \delta M_x + 2\gamma^2 H_\mathrm{an} H_{x} \\
&- \alpha \gamma (2H_\mathrm{ex} + H_\mathrm{an}) \partial_t \delta {M_x},
\end{aligned}
\end{equation}
where $\delta M_x = M_{1,x} + M_{2,x}$ and $\omega_\mathrm{sp} = \gamma \sqrt{H_\mathrm{an} ( 2H_\mathrm{ex} + H_\mathrm{an} )}$ is the eigenfrequency of the spin-wave, which is field independent \cite{Keffer1952,RezendeTutorial2019}. Under the assumption of harmonic oscillation $\delta M_x \propto e^{-i\omega t}$, we can derive the susceptibility of the system defined as $\chi \equiv \delta M_x / H_x$, where
\begin{equation}
\chi = \frac{2\gamma^2 H_\mathrm{an}M_s}{\omega_\mathrm{sp}^2 -i \alpha \gamma (2H_\mathrm{ex} + H_\mathrm{an}) -\omega^2}.
\end{equation}

On the other hand, the magnetization dynamics should also fulfil the Maxwell equations \cite{Yuanarxiv2024}
\begin{equation}\label{MaxwellHM}
(\nabla^2 +k^2) \mathbf{H}- \nabla(\nabla \cdot \mathbf{H}) + k^2\delta \mathbf{M} =0,
\end{equation}
where $k^2 = \epsilon \mu_0 \omega^2$ with $\epsilon$ being the permittivity of the medium and $\mu_0$ being the vacuum permeability. Substituting the relation $\delta M_x = \chi H_x$ into Eq. \eqref{MaxwellHM}, we have
\begin{equation}
(1+\chi)k_2^2 - (k_{2y}^2 + k_{2z}^2)=0
\end{equation}
where $\mathbf{k}_{i}$ represents the electromagnetic wavevector with $i=3,2$ for dielectric layer and AFM layer shown in Fig. \ref{fig1}, respectively.
In the long wavelength or magnetostatic limit, $1+ \chi=0$, the first term dominates the equation and thus one has $1+\chi=0$. Through straightforward mathematics, one can derive the resonance frequency with the revisions of dipolar fields \cite{YuanPRR2019} as $\omega_0 = \sqrt{\omega_\mathrm{sp}^2 + 2 \omega_\mathrm{an} \omega_m}$ in the absence of damping. Note that $\omega_\mathrm{sp}^2/(2 \omega_\mathrm{an} \omega_m) \sim H_\mathrm{ex}/M_s \gg 1$ for crystalline magnet, hence the renormalized resonance frequency is very close to the one without taking account of the dipolar interactions.

Furthermore, by solving the Maxwell equation $\mathbf{k}\times \mathbf{B} = -\omega \epsilon/c^2 \mathbf{E}$ with $\mathbf{B} = \mu_0(\mathbf{H} + \mathbf{M})$ and $c$ being the speed of light, we obtain the electric components of EM wave inside the AFM as
\begin{equation}
E_{2y}=-\frac{(1+\chi)k_{2z}}{\omega \epsilon_0 \epsilon_2} H_x,~E_{2z}=\frac{(1+\chi)k_{2y}}{\omega \epsilon_0 \epsilon_2} H_x,
\end{equation}
where $\epsilon_0$ is the vacuum permittivity. 

Similar to the ferromagnetic case \cite{Yuanarxiv2024}, the eigen-excitation of the hybrid system corresponds to the exponential decay of spin-wave modes in the magnetic layer and the evanescent EM waves in the dielectric layer. By solving the Maxwell equations, we can explicitly obtain the evanescent modes in the dielectric as
\begin{equation}
E_{3y}^{(+)} = -\frac{i\kappa_3}{\omega \epsilon_3 \epsilon_0} H_{3x}^{(+)},~E_{3z}^{(+)} = \frac{k_{3y}}{\omega \epsilon_3 \epsilon_0} H_{3x}^{(+)},
\end{equation} 
where the sign ($\pm$) means exponential increase(decrease) modes in the positive $z-$axis, $\kappa_3=\sqrt{k_{3y}^2-\omega^2\epsilon_3/c^2}$ is the decay coefficient of EM wave.

Now, we are ready to match the boundary conditions at the interface of dielectric and magnetic layers. The continuity of tangential components of magnetic and electric fields at the interface requires that
\begin{equation}\label{continumEH}
E_{3y}^{(+)}=E_{2y}^{(-)}, H_{3x}^{(+)}+\sigma E_{2y}^{(-)}=H_{2x}^{(-)},
\end{equation}
where the surface plasmon excitation in graphene is modelled as a surface electric current $j=\sigma E_{2y}^{(-)}$, where $\sigma$ is the ac conductivity of graphene. This approximation has been widely used to treat plasmons in graphene \cite{Mikhailov2007,Jablan2009}. It would be interesting to consider how to bridge this approach with the one based on electron-electron interaction in the future \cite{Grigorenko2012}. In THz regime, the graphene conductivity is well described by the Drude form $\sigma = \sigma_0 E_F/(\pi\Gamma - i\pi\hbar \omega)$ with $\sigma_0= e^2/4\hbar$, $E_F$ being the Fermi energy and $\Gamma$ being the relaxation rate of carriers.

\begin{figure}
	\centering
	\includegraphics[width=0.48\textwidth]{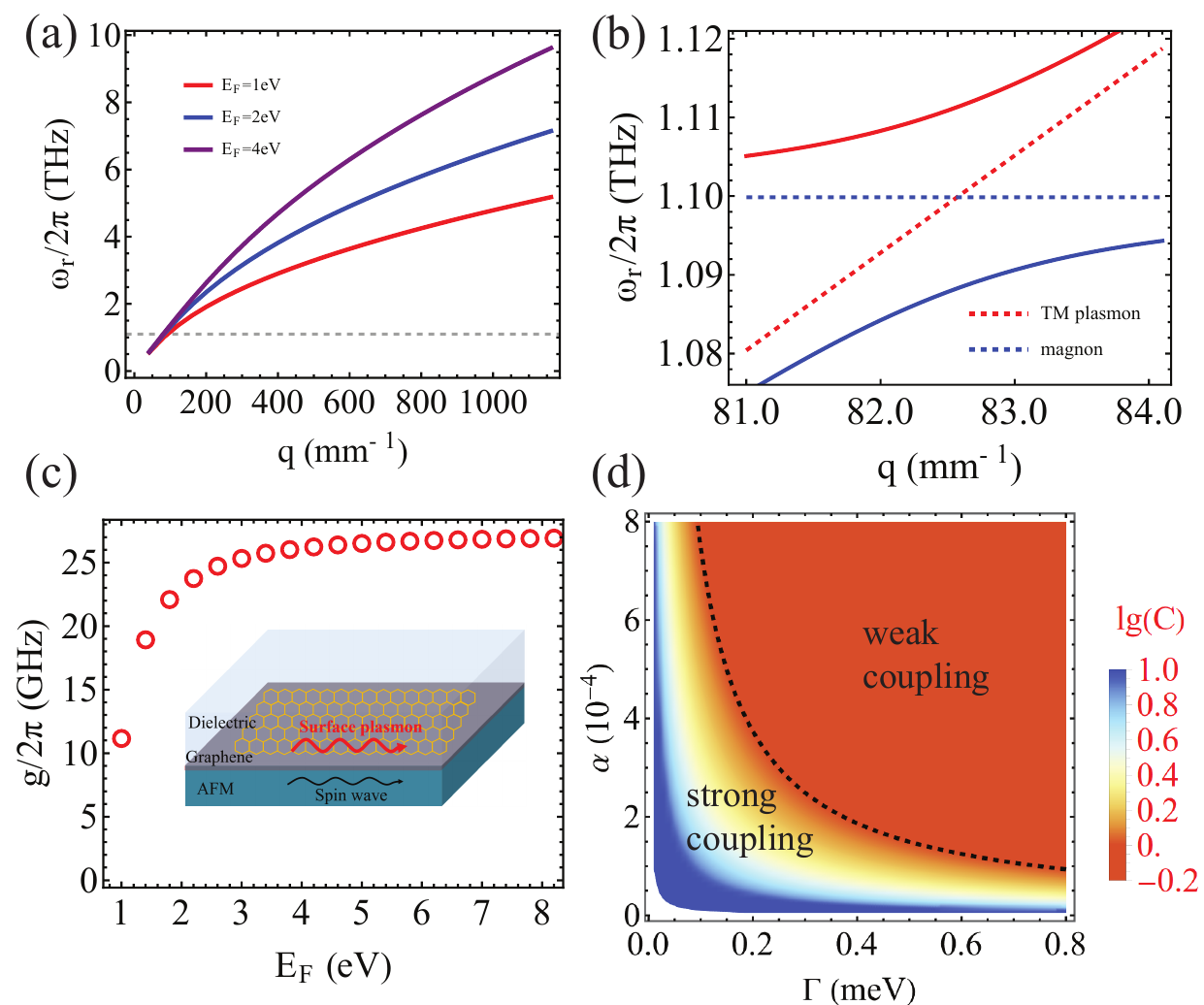}\\
	\caption{(a) Dispersion of surface plasmons in the DE/GRA/DE structure at different Fermi energy. The black dashed line indicates the frequency of antiferromagnetic magnons. $\epsilon_3=2,~\epsilon_2=11.9$ for NiO \cite{RaoJAP1965}. (b) Anticrossing spectrum between magnons and plasmons in the hybrid DE/GRA/AFM structure. The gap depth at resonance signals the coupling strength. $E_F=2$ eV. (c) Coupling spectrum between magnons and plasmons as a function of Fermi energy, which can be tuned by gating voltage. (d) Cooperativity in the phase diagram of $\alpha$ and $\Gamma$. $E_F=2$ eV. All the other parameters are the same as those in Fig. \ref{fig2}.}\label{fig3}
\end{figure}

To guarantee a nontrivial solution of the electromagnetic fields according to Eq. \eqref{continumEH}, it is required that
\begin{equation}\label{dispersion_eq_AFM}
-\frac{\kappa_3}{\omega \epsilon_3} \left ( i \frac{\sigma}{\epsilon_0} - \frac{\omega \epsilon_2}{\delta_p}  \right )=1,
\end{equation}
where $\delta_p$ ($p\in\{ \mathrm{AFM},\mathrm{DE}\}$) depends on the nature of medium 2 with the form $\delta_\mathrm{AFM} = -(1+\chi)\kappa_2$ and $\delta_\mathrm{DE} = -\kappa_2$.
For a dielectric substrate with $\delta_\mathrm{DE}$, we recover the known dispersion of surface plasmon in graphene \cite{Jablan2009}
\begin{equation}\label{dispersion_eq_DE}
\frac{4 \sigma_0 E_F}{\pi \epsilon_0 \hbar \omega^2} = \frac{\epsilon_2}{\sqrt{q^2 - \omega^2\epsilon_2/c^2}}+\frac{\epsilon_3}{\sqrt{q^2 - \omega^2\epsilon_3/c^2}},
\end{equation}
where we have imposed the requirement of momentum conservation at the interface $k_{2y} = k_{3y} \equiv q$. Note that this dispersion \eqref{dispersion_eq_DE} is quite different from that of the hybrid TE plasmon-magnon excitation \cite{Yuanarxiv2024}. In the electrostatic limit, $q \gg \omega \sqrt{\epsilon_i}/c$, Eq. \eqref{dispersion_eq_DE} can be analytically solved as
$\omega=\sqrt{8\sigma_0 E_F/(\pi \epsilon_0 \hbar (\epsilon_2 + \epsilon_3))q} \propto \sqrt{q}$, similar to the dispersion of surface plasmon on the metal surface \cite{Meierbook}. The tunability of Fermi energy $E_F$ in a 2D material readily allows for the tunability of the dispersion of surface plasmons, which is not present in normal metal and is a unique feature of 2D systems. Figure \ref{fig3}(a) shows that the dispersion relation at $E_F =1$ eV (red line), 2 eV (blue line), and 4 eV (purple line), respectively, by numerically solving Eq. \eqref{dispersion_eq_DE}.

When the substrate is an AFM, the surface spin wave in the THz regime will be excited and its frequency crosses the plasmon dispersion (black dashed line in Fig. \ref{fig3}(a). At the crossing point, the momentum and energy of surface magnons match that of the surface plasmon, and it is expected that a gap will open due to the spatial overlap of EM fields generated by plasmons and magnons inside the hybrid structure. To verify this point, we recall the dispersion relation \eqref{dispersion_eq_AFM} and simplify it as
\begin{equation}
\frac{4 \sigma_0 E_F}{\pi \epsilon_0 \hbar \omega^2} = \frac{\epsilon_2}{\sqrt{q^2 - \omega^2\epsilon_2/c^2}}+\frac{\epsilon_3}{(1+\chi)\sqrt{q^2 - (1+\chi)\omega^2\epsilon_3/c^2}}.
\end{equation}

Figure \ref{fig3}(b) shows the numerical solution of this equation at $E_F =2$ eV. A clear anticrossing feature can be identified at the resonance point $\omega_\mathrm{sp}/2\pi = 1.1~$THz. The gap of the anticrossing around 23 GHz characterizes the coupling strength between magnons and plasmons. Thanks to the tunability of the surface plasmon dispersion in 2D material by electric gating, the coupling strength between magnons and surface plasmons can also be tuned. Figure \ref{fig3}(c) shows that the coupling strength increases with the electron Fermi energy and saturates at a higher Fermi energy. This is because the dispersion of plasmon approaches $\omega=c/\sqrt{\epsilon}$ for larger $E_F$ ($\epsilon$ is the larger one of $\epsilon_2$ and $\epsilon_3$) according to Eq. \eqref{dispersion_eq_DE}, which is insensitive to the position of Fermi level. Then, the coupling behavior between the plasmon and magnon also becomes insensitive to the Fermi energy.

Taking account of the dissipation of magnons $\gamma_m = \alpha \omega_0$ and surface plasmons $\Gamma$, we can calculate the cooperativity $C=g^2/(\gamma_m \Gamma) $ \cite{Huebl2013} with $g$ being the coupling strength and show the result in Fig. \ref{fig3}(d). For typical values of magnetic damping $\alpha\sim 10^{-4}$ and electron relaxation rate $\Gamma \sim 0.5$ meV, the magnon and surface plasmon have already reached a strong coupling regime.


\section{Probe the coupling}
We have shown that antiferromagnetic magnons and surface plasmons can reach a strong coupling regime in a hybrid dielctric/graphene/antiferroamgnet structure. Here, we further show that such a coupling can be probed by measurement of the reflection spectrum of a THz wave from the hybrid structure. The idea is that the hybridized excitations of magnons and surface plasmons take away energy electromagnetic energy and result in double minimums in the reflection spectrum. Next, we will explicitly verify this point by analytically calculating the reflection spectrum.

\begin{figure}
	\centering
	\includegraphics[width=0.48\textwidth]{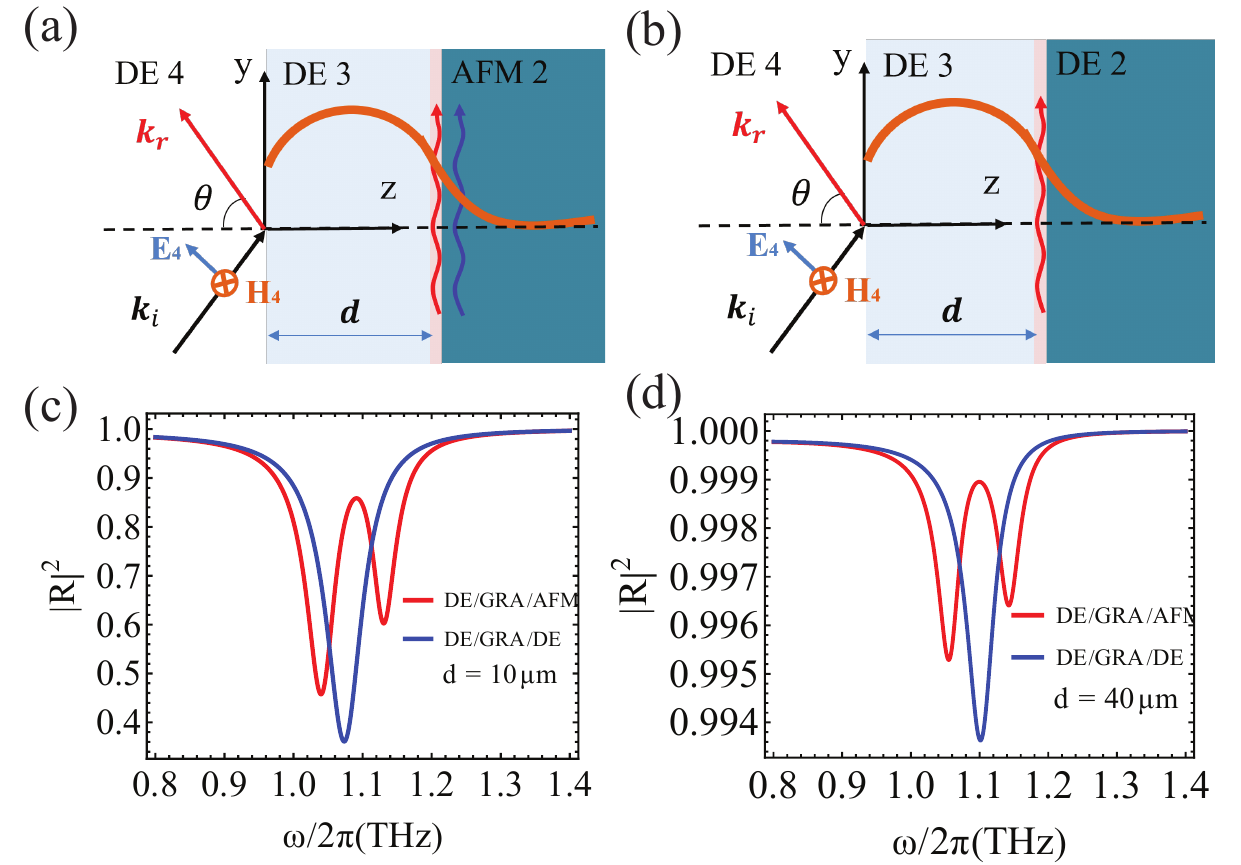}\\
	\caption{(a-b) Scheme of the two setups to measure the reflection rate of the hybrid DE/GRA/AFM and DE/GRA/DE structures. (c-d) Reflection rate of the hybrid system as a function of incident wave frequency for $d=10~\mathrm{\mu m}$ and $40~\mathrm{\mu m}$, respectively. Parameters are $E_F=2~\mathrm{eV},~\Gamma=0.1~\mathrm{meV},~\alpha=10^{-3},~\theta=39.4^\circ$.}\label{fig4}
\end{figure}

We add another input layer (dielectric medium 4) on top of the dielectric layer in Fig. \ref{fig1} and consider a hybrid DE/DE/GRA/AFM(DE) structure as shown in Fig. \ref{fig4}(a-b). The refractive index of medium 4 has to be larger than that of medium 3 ($\epsilon_4>\epsilon_3$) to guarantee that the incident electromagnetic wave can generate an evanescent wave in the medium 3 above a critical angle. By solving the Maxwell equations, it is straightforward to have the incident TM wave as
\begin{subequations}
\begin{align}
&\mathbf{H}_4^{(i,r)}= (H_{4x}^{(i,r)},0,0) e^{iqy+ik_{4z}^{(i,r)}z},\\
&\mathbf{E}_4^{(i,r)} = (0, E_{4y}^{(i,r)},E_{4z}^{(i,r)}) e^{iqy+k_{4z}^{(i,r)}z},
\end{align}
\end{subequations}
with 
\begin{equation}
E_{4y}^{(i,r)}= -\frac{k_{4z}^{(i,r)}}{\omega \epsilon_4 \epsilon_0} H_{4x}^{(i,r)},~E_{4z}^{(i,r)} = \frac{q}{\omega \epsilon_3 \epsilon_0} H_{4x}^{(i,r)},
\end{equation}
where the indices $i/r$ stand for the incident/reflected waves, and $q\equiv k_{4y}$ is the in-plane momentum of the electromagnetic wave, which should be conserved in the whole hybrid structure.
In the dielectric medium 3, now we have both exponential increase and decay modes due to its finite thickness ($d$). The full expression of magnetic and electric components is a linear superposition of these two modes, i.e.
\begin{subequations}
\begin{align}
&\mathbf{H}_3= (H_{3x}^{(+)},0,0) e^{iqy+\kappa_3 z} + (H_{3x}^{(-)},0,0) e^{iqy-\kappa_3z},\\
&\mathbf{E}_3 = (0, E_{3y}^{(+)},E_{3z}^{(+)}) e^{iqy+\kappa_3z} + (0, E_{3y}^{(-)},E_{3z}^{(-)}) e^{iqy-\kappa_3z}.
\end{align}
\end{subequations}
The continuity of the tangential components of magnetic and electric fields at the interfaces of media 4-3 ($z=0$) and media 3-2 ($z=d$) require that
\begin{subequations} \label{two-interfaces-bc}
\begin{align}
&H_{4x}^{(i)} + H_{4x}^{(r)} = H_{3x}^{(+)} + H_{3x}^{(-)},\\
&E_{4y}^{(i)} + E_{4y}^{(r)} = E_{3y}^{(+)} + E_{3y}^{(-)},\\
&H_{3x}^{(+)}e^{\kappa_3d} + H_{3x}^{(-)}e^{-\kappa_3d} + \sigma E_{2y}^{(-)}e^{-\kappa_2d}= H_{2x}^{(-)}e^{-\kappa_2d},\\
&E_{3y}^{(+)}e^{\kappa_3d} + E_{3y}^{(-)}e^{-\kappa_3d} = E_{2y}^{(-)} e^{-\kappa_2d}.
\end{align}
\end{subequations}
Here, we have shifted the coordinate center to the interface of media 4-3 for simplicity. By solving the linear set of equations \eqref{two-interfaces-bc}, we derive the reflection coefficient of the system as
\begin{equation}
R \equiv \frac{H_{4x}^{(r)}}{H_{4x}^{(i)}}=\frac{\kappa_3 \epsilon_4 \eta_1 + ik_z \epsilon_3 \eta_2}{\kappa_3 \epsilon_4 \eta_1 - ik_z \epsilon_3 \eta_2},
\end{equation}
where
\begin{equation}
\begin{aligned}
&\eta_1 \equiv (-i\delta_p \sigma + \epsilon_0\epsilon_2 \omega) \kappa_3 \sinh(\kappa_3d)- \epsilon_0 \epsilon_3 \delta_p \omega \cosh(\kappa_3 d),\\
&\eta_2 \equiv (-i\delta_p \sigma + \epsilon_0\epsilon_2 \omega) \kappa_3 \cosh(\kappa_3d)- \epsilon_0 \epsilon_3 \delta_p \omega \sinh(\kappa_3 d).\\
\end{aligned}
\end{equation}
Here $p=\mathrm{DE}$ for the DE/DE/GRA/DE structure and $p=\mathrm{AFM}$ for the DE/DE/GRA/AFM structure, as defined below Eq. \eqref{dispersion_eq_AFM}.

Figure \ref{fig4}(c) shows the reflection rate $|R|^2$ as a function of the incident wave frequency for $d=20~\mu$m. Clearly, two reflection minima are identified (red line), corresponding to the upper and lower branches of the hybrid modes in Fig. \ref{fig3}(b), respectively. As a comparison, only one dip structure is identified corresponding to the bare plasmon mode when the AFM layer is replaced by a standard dielectric layer, as shown in Fig. \ref{fig4}(c) (blue line). As the thickness of dielectric layer (medium 3) increases, the reflection rate is approaching 1, indicating that only tiny surface plasmons are excited at the graphene layer. This is because the electromagnetic wave decays exponentially in medium 3 and is not strong enough to excite the surface wave at the graphene after propagating a longer distance $d$. In the limit $d \rightarrow \infty$, $R=1$, we recover the textbook result of total reflection \cite{JacksonEMbook}.

\section{Plasmon assisted AFM-AFM coupling}

\begin{figure}
	\centering
	\includegraphics[width=0.48\textwidth]{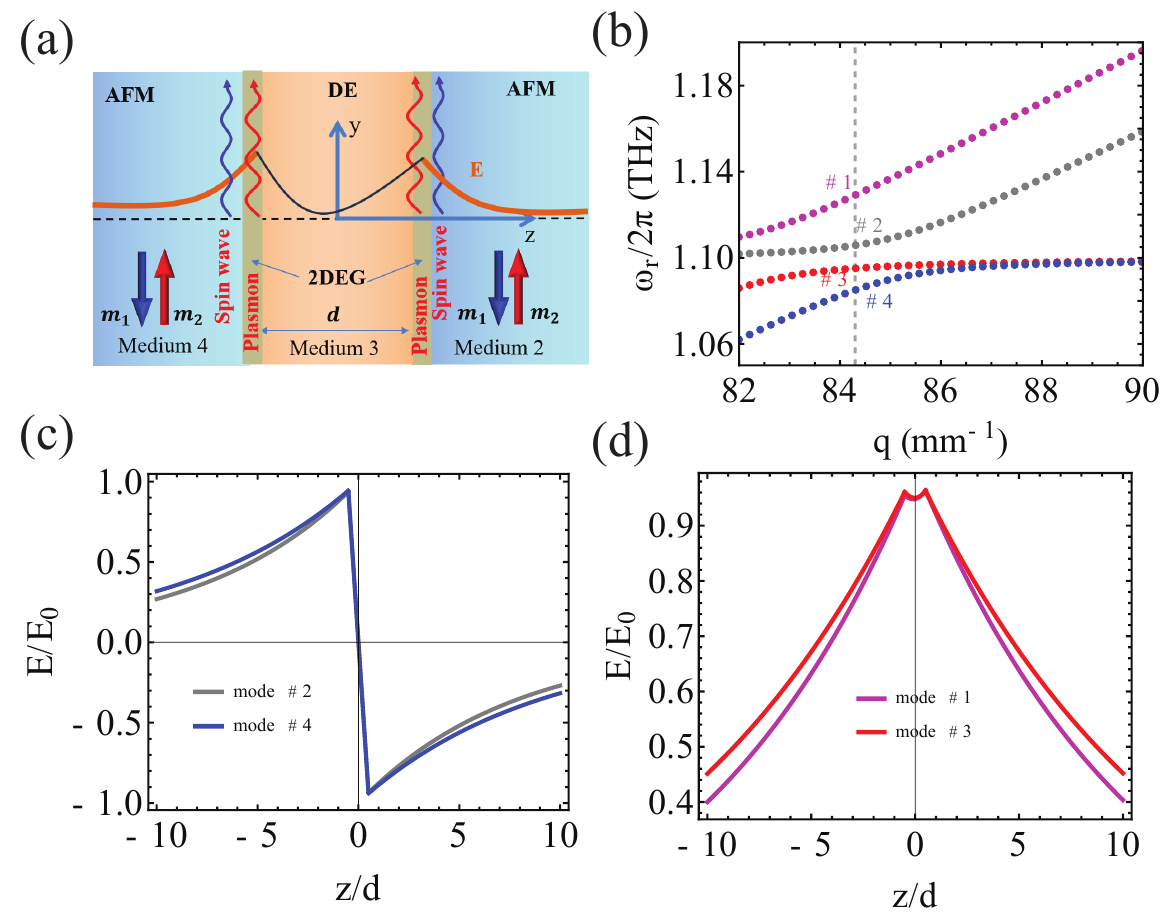}\\
	\caption{(a) Scheme of a hybrid AFM/GRA/DE/GRA/AFM structure. The surface plasmons excited at the graphene layers can mediate the coupling between two spatially separated antiferromagnets. (b) Dispersion relation of the hybrid modes calculated by numerically solving Eq. \eqref{symm-eq} (Magenta and red lines) and Eq. \eqref{antisymm-eq} (Gray and blue lines), respectively . Parameters are $d=4~\mathrm{\mu m}, \epsilon_4=\epsilon_2=11.9, \epsilon_3=2$. (c) and (d) are the spatial distribution of electric fields in the antisymmetric (modes $\#2$ and $\#4$) and symmetric modes (modes $\#1$ and $\#3$) at the crossing points between the vertical dashed line and the four hybrid modes shown in (b).}\label{fig5}
\end{figure}

As an application of the strong magnon-plasmon coupling, we show how the plasmon mode can mediate the coupling of magnon excitations in two well-separated antiferromagnets. We consider a hybrid AFM/GRA/DE/GRA/AFM structure as shown in Fig. \ref{fig5}. Following a similar approach to match the continuity of electric and magnetic fields at the interfaces of media 4-3 and 3-2, we find that the eigenexcitation of the hybrid system should satisfy the following equation

\begin{equation}\label{afm-afm-dispersion}
\begin{aligned}
&e^{-\kappa_3d} \left(\frac{\omega \epsilon_3}{\kappa_3} + \frac{\omega \epsilon_2}{\delta_{m2}} - \frac{i\sigma_2}{\epsilon_0}\right) \left(\frac{\omega \epsilon_3}{\kappa_3} - \frac{\omega \epsilon_4}{\delta_{m4}} - \frac{i\sigma_4}{\epsilon_0}\right)=\\
&e^{\kappa_3d} \left(\frac{\omega \epsilon_3}{\kappa_3} - \frac{\omega \epsilon_2}{\delta_{m2}} + \frac{i\sigma_2}{\epsilon_0}\right) \left(\frac{\omega \epsilon_3}{\kappa_3} + \frac{\omega \epsilon_4}{\delta_{m4}} + \frac{i\sigma_4}{\epsilon_0}\right).
\end{aligned}
\end{equation}
Here we have shifted the coordinate origin to the center of the middle dielectric layer with thickness $d$. The characterizing function $\delta_{mi}=-1(1+\chi_i)\kappa_i$, where $\chi_i$ and $\kappa_i$ are respectively the magnetic susceptibility and decay coefficients of the $i-$th layer.  

Without loss of generality, we assume that $\epsilon_2=\epsilon_4$, $\sigma_2 = \sigma_4$, then the dispersion equation can be factorized and analytically solved as
\begin{subequations}\label{two-hybrid-modes}
\begin{align}
\sinh \frac{\kappa_3 d}{2} \frac{\omega \epsilon_3}{\epsilon_3} + \cosh \frac{\kappa_3 d}{2} \left ( \frac{\omega \epsilon_2}{(1+\chi)\kappa_2} + \frac{i \sigma}{\epsilon_0} \right ) = 0, \label{symm-eq}\\
\cosh \frac{\kappa_3 d}{2} \frac{\omega \epsilon_3}{\epsilon_3} + \sinh \frac{\kappa_3 d}{2} \left ( \frac{\omega \epsilon_2}{(1+\chi)\kappa_2} + \frac{i \sigma}{\epsilon_0} \right ) = 0, \label{antisymm-eq}
\end{align}
\end{subequations}
where $\kappa_2 =\kappa_4 \equiv \kappa, \chi_2 =\chi_4 \equiv \chi$,
When $d \rightarrow \infty$, these two equations are identical to each other, whose solution recovers the two set of hybrid magnon-plasmon mode as already presented in Fig. \ref{fig3}.  

For a dielectric layer with finite thickness, the hybrid magnon-plasmon mode at the left region ($z<0$) will overlap with that in the right region ($z>0$) through the middle dielectric layer. Such a hybridization will result in two new classes of hybrid modes, as shown in Fig. \ref{fig5}(b).  Here the black and red lines refer to the solutions of Eqs. \eqref{symm-eq} and \eqref{antisymm-eq}, which correspond to the symmetric and antisymmetric mode, respectively. 
To clarify, we can explicitly solve the linear set of equations characterizing the boundary conditions and derive the spatial distribution of electric fields across the hybrid structures. Figures \ref{fig5}(c) and (d) show that symmetric and antisymmetric modes are identified for the mode satisfying  Eqs. \eqref{symm-eq} and \eqref{antisymm-eq}, respectively. Such an indirect coupling channel share certain similarities with the coupling between two magnons mediated by phonons in a nonmagnetic insulator, where both in-phase and out-of-phase motions can be generated \cite{AnPRB2020}.


\section{Discussion and conclusion}

First, all our previous calculations focus on the monolayer graphene. For a double-layer graphene attached on top of an antiferromagnet, the modelling technique is quite similar to the one shown in Fig. \ref{fig5}, where a dielectric layer replaces the medium 4. The magnon-plasmon coupling spectrum of the hybrid structure can still be routinely described by Eq. \eqref{afm-afm-dispersion} after replacing $\delta_{m4}$ by its dielectric counterpart $-\kappa_4$. When the interlayer distance of the two grpahene layers is very small that $\kappa_3 d\ll 1$, we can analytically solve the dispersion relation as
\begin{equation}\label{dispersion_eq_2Gra}
-\frac{\kappa_3}{\omega \epsilon_3} \left ( i \frac{2\sigma}{\epsilon_0} - \frac{\omega \epsilon_2}{\delta_\mathrm{AFM}}  \right )=1,
\end{equation}
where we have assumed $\kappa_4=\kappa_3$ for simplicity. Compared with the monolayer case derived in Eq. \eqref{dispersion_eq_AFM}, the graphene conductivity is doubled. Since the conductivity of graphene is proportional to the electron Fermi energy, the Fermi energy will then be shifted upwards, and hence the coupling strength between magnons and plasmons will become stronger, as discussed in Fig. \ref{fig3}(c). Physically, a larger electron Fermi energy will make the bare plasmon dispersion more steep (Fig. \ref{fig3}(a)) and crosses the frequency of spin wave at a smaller wavevector $q$. Then the decay length of surface plasmon toward the dielectric medium 3 ($1/\kappa = 1/\sqrt{q^2 - \omega^2 \epsilon_3/c^2}$) will become larger, resulting in a stronger overlap between plasmon mode and surface spin wave mode. This is why the effective coupling between magnons and plasmons becomes stronger.

Further, the above analysis on double-layer graphene may not work for bilayer graphene with atomic interplay distance close to that in graphite ($\sim 3.4$ Å). Then, the conductivity of the system will be modified by the van der Waals interaction between the two layers. Take the AA-stacked bilayer graphene as an example, the two copies of band structure of monolayer graphene will shift upward and downward by $\Delta$, respectively, with $\Delta$ being the interlayer coupling strength. The resulting dynamic conductivity is \cite{TabertPRB2012}
\begin{equation}
\begin{aligned}
\sigma(\omega) =&\frac{8i\sigma_0}{\pi \hbar \omega} \max(\Delta, E_F) + \frac{i}{\pi} \ln \frac{\hbar \omega - 2 |\Delta - E_F|}{\hbar \omega + 2 |\Delta - E_F|} \\
&+  \frac{i}{\pi} \ln \frac{\hbar \omega - 2 |\Delta + E_F|}{\hbar \omega + 2 |\Delta + E_F|}.
\end{aligned}
\end{equation}
When the Fermi energy is much larger than the interlayer coupling, i.e. $E_F \gg \Delta$, the conductivity is reduced to $\sigma = 8i\sigma_0/(\pi\hbar \omega)$, which is exactly two times of the conductivity of monolayer graphene. Then we recover the results of double-layer graphene. When the Fermi energy is comparable to the interlayer coupling, especially when their energy difference $|E_F-\Delta|$ is close to the magnon energy, the results can be very different. A detailed discussion on the bilayer case will be given elsewhere.

Lastly, we would like to discuss a few literature on the coupling between magnons and plasmons. Costa et al. \cite{CostaNL2023} reported strong coupling between surface plasmon and magnons in graphene-2D ferromagnetic heterostructures by assuming an artificial Zeeman interaction between plasmonic magnetic fields and spins. Here we notice that one has to be very careful to match the boundary conditions at the interface between graphene and magnets. Also, generating THz uniform magnon mode in a ferromagnet is very challenging due to the influence of exchange interaction. Dyrdal et al. \cite{DyrdalPRB2023} reported the hybridization of magnons and plasmon by spin-orbit interaction in 2D magnets. Ghosh et al \cite{GhoshPRB2023} showed that plasmons can hybridize with weakly dispersive optical magnons in 2D honeycomb magnets. A proof-of-concept experiment on this coupling is still lacking. Recently, Xiong et al. \cite{Xiong2024} observed the strong coupling between magnons and spoof surface plasmons in a hybrid structure of magnetic sphere and microwave spiral resonator in experiments.

In conclusion, we have shown that surface plasmons in 2D material can reach strong coupling with the surface magnons in antiferromagnets at the THz regime. The coupling strength is tunable by electric gating of the 2D material, where the Fermi energy of electrons and thus the conductivity can be changed. To probe the coupling, we propose measuring the reflection spectrum of the hybrid system, where the hybrid magnon-plasmon modes take away electromagnetic energy and generate two reflection minima. We further show that the plasmons can mediate the coupling between two AFMs over several micrometers. Both symmetric and antisymmetric hybrid modes can be generated. Our finding may open a novel platform to study the interplay of magnon spintronics and plasmonics, where antiferromagnets may manifest their THz dynamics in particular. Further, it would be interesting to dive into the quantum regime by quantizing the classical field profiles and to study the quantum correlations between magnons and plasmons, potentially including plasmons to the family of quantum magnonics \cite{YuanQM}. This will be the focus of our follow-up work.

{\it Acknowledgments.}---This work was supported by the National Key R$\&$D Program of China (2022YFA1402700) and the Dutch Research Council (NWO).

{}
\end{document}